# 10 THz Ultrafast Function Generator
# - generation of rectangular and triangular pulse trains-

**Kanaka Raju Pandiri and Masayuki Katsuragawa***

*Department of Engineering Science, University of Electro-Communications, 1-5-1 Chofugaoka, Chofu-Shi, Tokyo, Japan 182-8585*
*\*katsura@pc.uec.ac.jp*

**Abstract:** We report the synthesis of arbitrary optical waveforms by manipulating the spectral phases of Raman sidebands with a wide frequency spacing line-by-line. Trains of rectangular and triangular pulses are stably produced at an ultrahigh repetition rate of 10.6229 THz, reminiscent of an ultrafast function generator.

**References and links**

1. A. Assion, T. Baumert, M. Brixner, B. Kiefer, V. Seyfried, M. Strehle, and G. Gerber, "Control of chemical reactions by feedback-optimized phase-shaped femtosecond laser pulses," Science **282**, 919–922 (1998).
2. R. J. Levis, G. M. Menkir, and H. Rabitz, "Selective bond dissociation and rearrangement with optimally tailored, strong-field laser pulses," Science **292**, 709–713 (2001).
3. K. Takiguchi, K. Okamoto, T. Kominato, H. Takahashi, and T. Shibata, "Flexible pulse waveform generation using silica-waveguide-based spectrum synthesis circuit," Electron. Lett. **40**, 537–538 (2004).
4. Z. Jiang, C.-B. Huang, D. E. Leaird, and A. M. Weiner, "Optical arbitrary waveform processing of more than 100 spectral comb lines," Nat. Photonics **1**, 463–467 (2007).
5. N. K. Fontaine, R. P. Scott, L. Zhou, F. M. Soares, J. P. Heritage, and S. J. B. Yoo, "Real-time full-field arbitrary optical waveform measurement," Nat. Photonics **4**, 248–254 (2010).
6. Z. Jiang, D.E. Leaird, and A.M. Weiner, "Line-by-line pulse shaping control for optical arbitrary waveform generation," Opt. Express **13**, 10431–10439 (2005).
7. S. E. Harris, and A. V. Sokolov, "Broadband spectral generation with refractive index control," Phys. Rev. A **55**, R4019–R4022 (1997).
8. A. V. Sokolov, D. R. Walker, D. D. Yavuz, G. Y. Yin, and S. E. Harris, "Raman generation by phased and antiphased molecular states," Phys. Rev. Lett. **85**, 562–565 (2000).
9. J. Q. Liang, M. Katsuragawa, F. L. Kien, and K. Hakuta, "Sideband generation using strongly driven Raman coherence in solid hydrogen," Phys. Rev. Lett. **85**, 2474–2477 (2000).
10. D. D. Yavuz, D. R. Walker, M. Y. Shverdin, G. Y. Yin, and S. E. Harris, " Quasiperiodic Raman technique for ultrashort pulse generation," Phys. Rev. Lett. **91**, 233602-223605 (2003).
11. S. E. Harris and A. V. Sokolov, "Subfemtosecond pulse generation by molecular modulation," Phys. Rev. Lett. **81**, 2894-2897 (1998).
12. A. V. Sokolov, D. R. Walker, D. D. Yavuz, G. Y. Yin, and S. E. Harris, "Femtosecond light source for phase-controlled multiphoton ionization," Phys. Rev. Lett. **87,** 033402-033405 (2001).
13. M. Y. Shverdin, D. R. Walker, D. D. Yavuz, G. Y. Yin, and S. E. Harris, "Generation of single-cycle optical pulse," Phys. Rev. Lett. **94**, 033904–033907 (2005).
14. M. Katsuragawa, K. Yokohama, T. Onose, and K. Misawa, "Generation of a 10.6-THz ultrahigh-repetition-rate train by synthesizing phase-coherent Raman-sidebands," Opt. Express **13**, 5628–5634 (2005).
15. W. J. Chen, Z. M. Hsieh, S. W. Huang, H. Y. Su, C. J. Lai, T. T. Tang, C. H. Lin, C. K. Lee, R. P. Pan, C. L. Pan, and A. H. Kung, "Sub-single-cycle optical pulse train with constant carrier envelope phase," Phys. Rev. Lett. **100**, 163906 (2008).
16. T. Suzuki, M. Hirai, and M. Katsuragawa, "Octave-spanning Raman comb with carrier envelope offset control," Phys. Rev. Lett. **101**, 243602 (2008).
17. S. N. Goda, M. Y. Shverdin, D. R. Walker, and S. E. Harris, "Measurement of Fourier-synthesized optical waveforms," Opt. Lett. **30**, 1222–1224 (2005).
18. Z. M. Hsieh, C. J. Lai, H. S. Chan, S. Y. Wu, C. K. Lee, W. J. Chen, C. L. Pan, F. G. Yee, and A. H. Kung, "Controlling the carrier-envelope phase of Raman-generated periodic waveforms," Phys. Rev. Lett. **102**, 213902 (2009).
19. Kanaka Raju P., T. Suzuki, A. Suda, K. Midorikawa, and M. Katsuragawa, "Line-by-line control of 10-THz-frequency-spacing Raman sidebands," Opt. Express **18**, 732–739 (2010).
20. S. J. Russell and P. Norvig, *Artificial Intelligence A Modern Approach* (Pearson Education, Inc., New Jersey, 2010), Chap. 4.

21. M. Katsuragawa and T. Onose, "Dual-wavelength injection-locked pulsed laser," Opt. Lett. **30**, 2421–2423 (2005).
22. T. Onose and M. Katsuragawa, "Dual-wavelength injection-locked pulsed laser with highly predictable performance," Opt. Express. **15**, 1600–1605 (2007).
23. A. Suda, Y. Oishi, K. Nagasaka, P. Wang, and K. Midorikawa, "A spatial light modulator based on fused-silica plates for adaptive feedback control of intense femtosecond laser pulses," Opt. Express **9**, 2–6 (2001).
24. T. Suzuki, N. Sawayama, and M. Katsuragawa, "Spectral phase measurement for broad Raman sidebands by using spectral interferometry," Opt. Lett. **33**, 2809–2811 (2008).

## 1. Introduction

The generation and measurement of arbitrary optical waveforms (AOW) have had a significant impact in relation to the coherent control of chemical reactions [1, 2], and optical communications [3–5]. Line-by-line control, in which spectral lines are resolved and manipulated individually, leads to a fundamentally new regime for AOW generation [4, 6]. Adiabatically generated Raman sidebands (Raman comb) [7-10] is suitable source for implementing line-by-line control since they have discrete characteristics typically a greater than THz frequency spacing and good mutual coherence. By using such Raman sidebands, a synthesized AOW can have an ultrahigh repetition rate corresponding to the wide frequency spacing of the Raman sidebands. Here, we report the synthesis of 10-THz ultrahigh repetition rate AOW using rotational Raman sidebands ($J = 2 \leftarrow 0$) in parahydrogen. We show that trains of rectangular and triangular pulses are stably produced, similar to an ultrafast function generator.

Before proceeding, we briefly review previous studies on the adiabatic Raman technique in relation to ultrafast technology. The potential for generating single cycle pulses was discussed theoretically [11] and shown experimentally [12, 13]. The generation of high-quality ultrashort pulses was reported based on a direct time-domain measurement [14]. The control of the carrier-envelope offset frequency was demonstrated with a cross correlation measurement [15] and the *f-2f* self-referencing technique [16]. Arbitrary optical amplitude waveform generation was discussed theoretically [17]. Carrier-envelope phase control was realized by developing a build up scheme of Raman sidebands [18]. Recently, a train of Fourier-transform-limited pulses was generated by a glass-plate-based line-by-line phase controller [19].

## 2. Numerical simulation: production of effective target for AOW generation

The AOW generation was performed in two stages. The first stage was numerical AOW generation on a computer. We employed the same spectrum as that in the actual AOW generation using Raman sidebands, and determined the optimal spectral phase for a target waveform. In the second stage, we generated Raman sidebands experimentally and synthesized the AOW by employing the numerically obtained optimal spectral phase as an effective target. In this section, we describe the process in the first stage for determining the optimal spectral phases (effective target) on a computer.

The gray bars in Fig. 1a represent seven discrete frequency components, designated $\Omega_{-4}$ to $\Omega_2$. The spectral intensity distribution and the frequency spacing (10.6229 THz) were set at the same as those employed for the actual AOW generation using the Raman sidebands. (Experimental details are provided in Sec. 3.) The numerical AOW generation was carried out by manipulating the spectral phases of these seven frequency components iteratively based on the hill climbing algorithm [20] to realize a target waveform. In this process, we defined a cost function, *C*, as the square of deviation of a shaped pulse from a target waveform, which is expressed as

$$C = \int \left| I^{tar}(t) - I(t) \right|^2 dt \tag{1}.$$

$I^{tar}(t)$ and $I(t)$ are the target and recurring waveforms in the time domain, respectively. The iteration process was initiated from a waveform in which all the spectral phases were zero. Then, the cost function, $C$, was minimized by changing the phase of each sideband line-by-line based on the hill climbing algorithm. A new phase change was accepted if the new cost function was smaller than that of the last accepted phase change. The iteration was stopped automatically when the cost function was saturated. The optimal spectral phases (effective target spectral phase) and the corresponding waveform (effective target waveform) were thus identified for the ideal target waveform, $I^{tar}(t)$.

A typical example when the target waveform, $I^{tar}(t)$, was set to an ultrahigh rep. rate train of the rectangular pulses (gray line in Fig. 1b; one cycle time: 94.1363 fs (10.6229 THz), duty cycle: 50%) is shown by the blue line with dots (effective target spectral phase) in Fig. 1a, and the blue curve (effective target waveform) in Fig. 1b. The distortion of the effective target waveform from the flat top of the ideal target waveform is attributed to the fundamental limitation imposed by the number (seven) of frequency components, which was restricted by the detection sensitivity in the present experimental system.

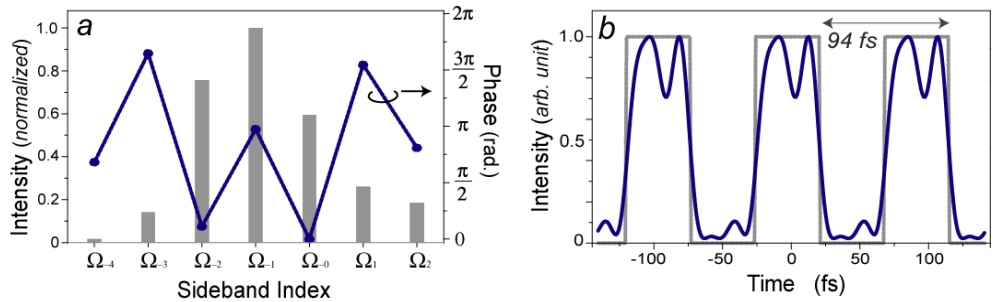

Fig.1 Production of an effective target for AOW generation. a, Sideband power spectrum (gray bars) and the generated optimal spectral phases (blue line and dots). b, Ideal rectangular pulse train (gray line) and the numerically obtained effective target waveform (blue line).

## 3. Experimental

The optimal spectral phases obtained numerically were employed as the effective target in the actual AOW generation stage using the Raman sidebands. In this section, we describe the details of the experimental configuration and the procedure for convergence with the effective target in the experimental system.

*3.1 Detailed configuration of experimental system*

The entire system consists of three parts that generate Raman sidebands, manipulate their spectral phases, and measure the spectral phases to produce the obtained waveform together with an error signal for feedback.

The Raman sidebands were generated by adiabatically driving the rotational Raman process ($J = 2 \leftarrow 0$) in parahydrogen [14, 16]. The density and temperature of the parahydrogen were set at $3 \times 10^{20}$ cm$^{-3}$ (interaction length: 15 cm) and 77K (liquid-nitrogen temperature), respectively. The two pump laser radiations, $\Omega_0$ and $\Omega_{-1}$, were produced with a dual-frequency injection-locked nanosecond pulsed Ti:sapphire laser [21, 22]. They were both Fourier-transform limited 6-nanosecond pulses (spectral width: ~ 50 MHz [21]) and they overlapped each other temporally and spatially. The frequencies of $\Omega_0$ and $\Omega_{-1}$ were set at 382.4280 THz (783.9186 nm) and 371.8050 THz (806.3162 nm), respectively, and were slightly detuned by $\delta = + 990$ MHz from the Raman resonance to satisfy the adiabatic condition [7, 14]. The two pump laser beams were then loosely focused in the parahydrogen with an $f$ = 800 mm lens. The peak intensity was estimated to be 5.7 GW / cm$^2$ (total pulse energy: 5.4 mJ).

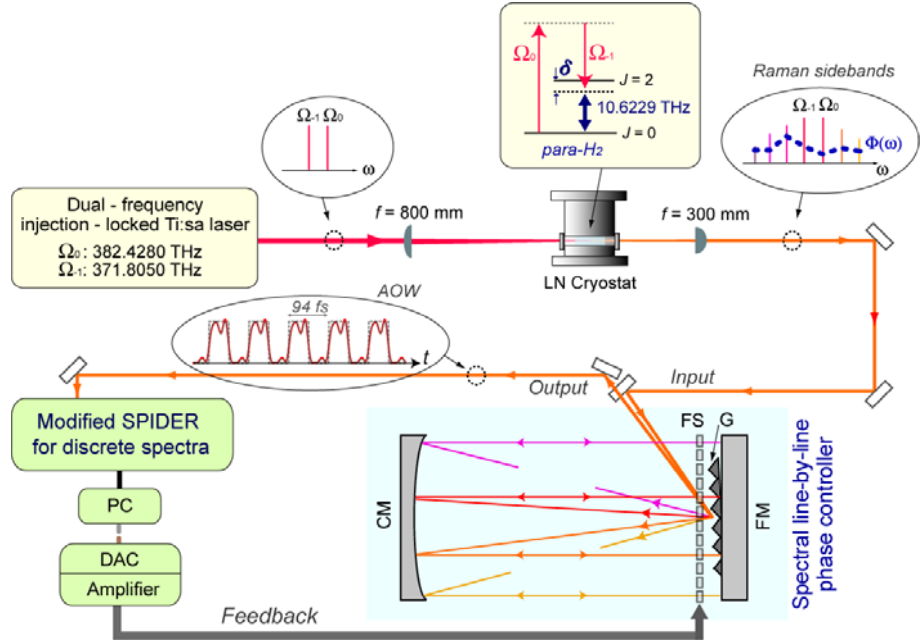

Fig.2. Experimental setup. LN Cryostat, liquid nitrogen cryostat; G, grating; FS, fused silica plates; FM, folding mirror; CM, concave mirror; PC, personal computer; DAC, digital–analog converter; AOW, arbitrary optical waveform.

The sidebands were generated in the forward direction and then introduced into the spectral line-by-line phase controller (SLLPC) after being collimated with an $f$ = 300 mm lens. The SLLPC was a spatial phase modulator consisting of 48 fused-silica plates. The system was configured in a folded 4$f$-configuration [23]. The incident Raman sidebands were dispersed with a grating (600 grooves / mm) and each single sideband was focused onto a single glass plate placed in the Fourier plane by using a concave mirror ($f$ = 0.5 m). The change in the spectral phase of each sideband was accomplished by changing the angle of the corresponding fused-silica plate through a bimorphous-type piezo actuator. The focused beams were reflected by the folding mirror positioned just behind the glass plates, and all the sidebands were finally recombined into one beam in space.

The phase controlled Raman sidebands and the corresponding waveform in the time domain were characterized by using modified spectral-phase interferometry for a direct electric-field reconstruction system for discrete spectra (SPIDER-DS) [24]. This system identified the relative spectral phases among the sidebands on the basis of the interference with a pair of sum frequencies, which were produced from the Raman sidebands and the two-frequency pump beams [24]. The SPIDER-DS system operated rapidly and generated an error signal (the deviation from the effective target spectral phase) on a LabVIEW based personal computer. Finally, the error signal was fed back to the SLLPC system through a set of 48-channel digital-to-analog converters and amplifiers (maximum applicable voltage: ± 60 V).

*3.2 Process of convergence with effective target*

The concrete process for convergence with an effective target spectral phase was as follows. First, we measured the sideband phase change caused by an applied voltage to a bimorphous-type piezo actuator in the SLLPC, as a prerequisite of quantitative phase manipulation. We scanned the applicable voltage, $\upsilon_n$, to the actuator over the full range (60 to - 60 V; minimum step: 1 V), to give the periodic phase change, $\Phi_n(\upsilon_n)$, at the n-th sideband. This periodic phase change also induced a periodic intensity change in the related interfered sum-frequency, $I_{n,n+1}$, in the SPIDER-DS [19]. We normalized this periodic intensity change to $-1$ to $1$ (designated by $\overline{I_{n,n+1}}$), and then reduced the relation: the relative spectral phase, $\Phi_{n+1}(\upsilon_{n+1}) - \Phi_n(\upsilon_n) = \cos^{-1}(\overline{I_{n,n+1}})$ versus the applied voltage, $\upsilon_n$, where $\upsilon_{n+1}$ was fixed. The same procedure was executed line-by-line for all the sidebands.

We can employ the above relation to directly set the spectral phases of the sidebands to obtain an effective target. This process, however, provided only coarse adjustment, since the SLLPC system included the hysteresis inherent in the piezo actuator. To ensure that the spectral phases converged sufficiently with the effective target, the process was followed by iterative phase manipulation based on the precise phase measurement provided by the SPIDER–DS. We produced the error signal from the SPIDER–DS measurement together with the above relation, fed this error signal back to the SLLPC, and then evaluated the spectral phases again with the SPIDER–DS. We repeated this set of procedures until the spectral phases of the sidebands reached the effective target within the minimum resolution of the SLLPC ($\pm$ 0.1 radians).

## 4. Results

The AOW generated using the Raman sidebands were examined for two cases: an ultrahigh rep. rate train of rectangular pulses and an ultrahigh rep. rate train of triangular pulses. We show the results in this section.

Figure 3 shows the spectrum of the generated Raman sidebands. Twelve sidebands from $\Omega_{-5}$ (329.314 THz) to $\Omega_6$ (446.165 THz) are seen clearly. From these twelve, we selected seven from $\Omega_{-4}$ (339.936 THz) to $\Omega_2$ (403.674 THz): highlighted by the yellow box, to use for the AOW generation.

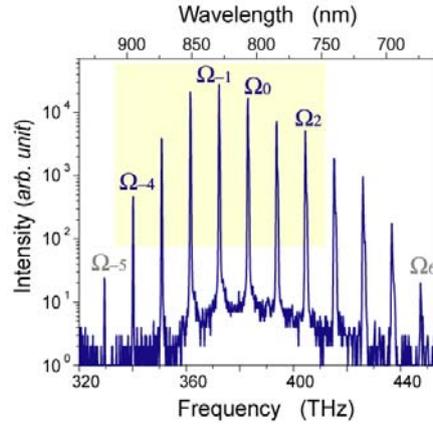

Fig. 3. Raman sideband spectrum. The seven sidebands from $\Omega_{-4}$ to $\Omega_2$ (highlighted by yellow box) were employed for AOW generation.

Figures 4a, b, and c show the results when the target waveform was set as a train of rectangular pulses (gray line in Fig. 4c; duty ratio: 50%). The initial spectral phases of the Raman sidebands are shown by the black dotted line in Fig. 4a, together with their power spectrum on a linear scale (gray bars: identical to those in Fig. 3). The blue line in Fig. 4a is the effective target spectral phases obtained numerically in advance as described in Sec. 2 and the blue curve in Fig. 4c is the corresponding effective target waveform in the time domain. We employed the process for converging the spectral phases of sidebands with the effective target and realized the spectral phases shown by the red line in Fig. 4a at the end of the iterative phase manipulation. The obtained spectral phases precisely matched the effective target (blue line).

An example of this convergence process is shown for the $\Omega_1$ sideband in Fig. 4b. Iteration number 0 indicates the initial phase. At the first coarse adjustment, which is realized by directly applying the relation of the relative spectral phase versus the applied voltage, this phase reached the value at the iteration number 1. Then, the adjustments based on the precise phase measurement by the SPIDER-DS were applied iteratively, and the phase (red square) was finally set at the target (blue line) within ± 0.1 radians after six successive trials.

The red curve in Fig. 4c shows the waveform realized in the time domain. A precise match with the effective target waveform (blue curve) was also found in the time domain. Note that this waveform was reduced for measuring the spectral phases of the sidebands with nanosecond pulsed envelopes, and they showed fine stability (< ± 0.1 radians) for several hours. This shows that the ultrahigh repetition rate (10.6229 THz) train of the rectangular waveform was stably produced with more than 10, 000 successive pulses.

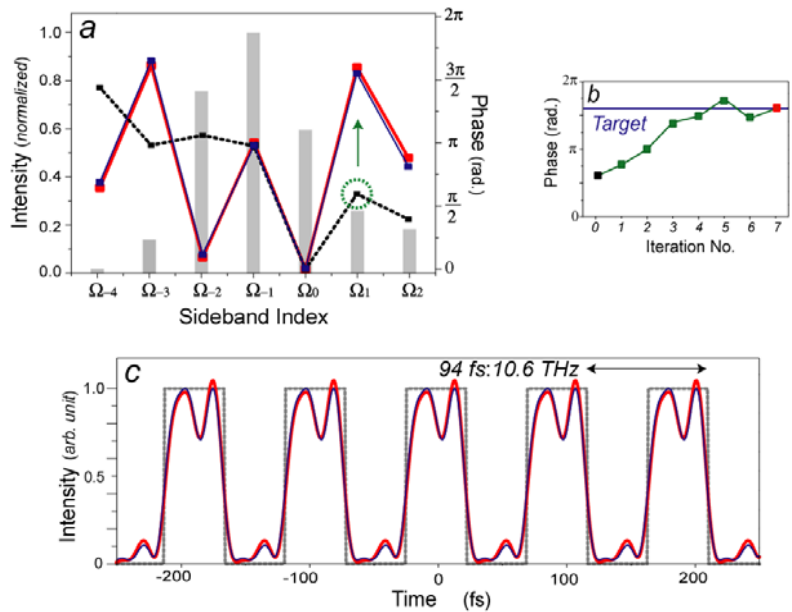

Fig. 4. Generation of an ultrahigh repetition rate train of rectangular pulses. *a*, Spectral phases (blue line: effective target, black dotted line: initial phase, red line: achieved phase) and power spectrum (gray bars) of the Raman sidebands. *b*, An example of the process that converges the spectral phase with the target (blue line), shown for the $\Omega_1$ sideband. *c*, A train of rectangular pulses: target waveform (gray line), effective target waveform (blue line), and the achieved waveform (red line).

Next we examined the case where the target was set at a train of triangular pulses. The results we obtained are shown in Fig. 5a, b and c, in a similar way to those in Fig. 4. AOW generation was also successfully achieved for this train of triangular pulses at an ultrahigh repetition rate of 10.6229 THz.

Here, one of the sideband phases, $\Omega_2$, was ~ 0.3 radians from the effective target. As described above, the present system offers the ability to control the spectral phases with ± 0.1 radian resolution. However, in some cases, we encountered degraded convergence behavior for a spectral phase with an unknown cause. This was such a case. However, we could confirm that here the obtained waveform matched the effective target waveform sufficiently well at this stage, and thereby halted the iterative phase manipulation.

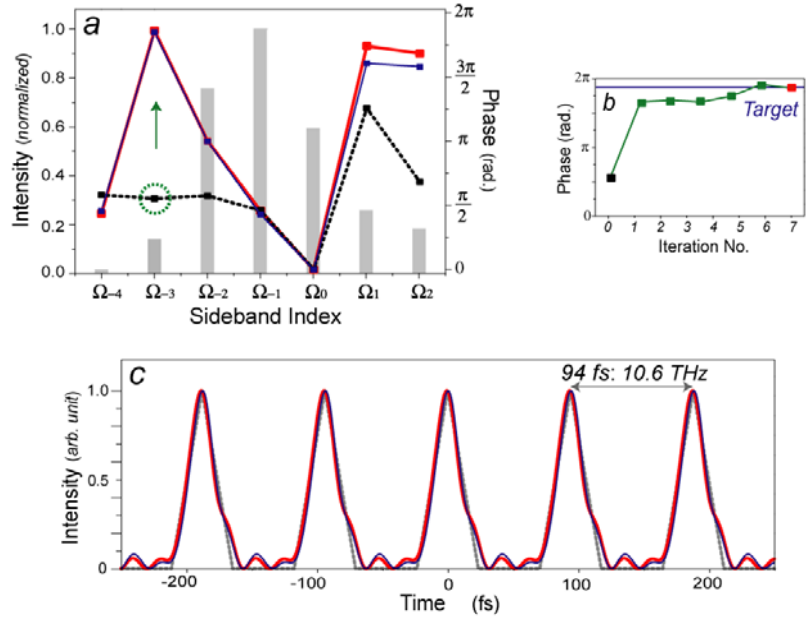

Fig. 5. Generation of an ultrahigh repetition rate train of triangular pulses. *a*, Spectral phases (blue line: effective target, black dotted line: initial phase, red line: achieved phase) and power spectrum (gray bars) of the Raman sidebands. *b*, An example of the convergence process of the spectral phase to the target (blue line), shown for the $\Omega_{-3}$ sideband. *c*, A train of triangular pulses: target waveform (gray line), effective target waveform (blue line), and the achieved waveform (red line).

### 5. Conclusion

We have described the synthesis of AOW using Raman sidebands with a wide frequency spacing, which were generated by adiabatically driving the rotational Raman process ($J = 2 \leftarrow 0$) in parahydrogen. Trains of rectangular and triangular pulses were produced at an ultrahigh repetition rate of 10.6229 THz as if the system were an ultrafast function generator. The trains were stably constituted over a nanosecond time scale with more than 10,000 shaped pulses.

Note that in the frequency region beyond 1 THz, various elementary excitations, such as phonon, magnon, superconductor gap, and so on, emerge. The key idea for the potential applications of the present ultrahigh-repetition-rate AOW light source would be manipulations of light-matter interactions in the new regime: on the basis of a resonance between not light-frequency but an ultrahigh-repetition-rate and characteristic frequencies of such elementary excitations.

### Acknowledgements

We are grateful to T. Suzuki for valuable advice. We thank K. Midorikawa, and A. Suda for useful discussions. We also thank N. Yamazaki, N. Sawayama, and A. J. Kiran for their technical help with the experiment.